\documentstyle[a4,12pt,epsf]{article}
\newcommand{\noi}{\noindent}
\newcommand{\eq}{\begin{equation}}
\newcommand{\en}{\end{equation}}
\newcommand{\eqa}{\begin{eqnarray}}
\newcommand{\ena}{\end{eqnarray}}

\newcommand{\vp}{{\vec p}}

\newcommand{\vx}{{\vec x}}

\newcommand{\bpartial}{{\bar \partial}}

\newcommand{\aleq}{\mbox{}_{\textstyle \sim}^{\textstyle < }}
\newcommand{\ageq}{\mbox{}_{\textstyle \sim}^{\textstyle > }}

\newcommand{\lra}{\longrightarrow}

\begin{document}

\renewcommand{\theequation}{\arabic{section}.\arabic{equation}}
\renewcommand{\thesection}{\arabic{section}}

\vspace{-2cm}

\hbox{}
\noindent March 1999  \hfill JINR E2--99--44

                      \hfill HUB--EP--99/09

\vspace*{1.0cm}

\begin{center}

{\Large Lorentz gauge and Gribov ambiguity} 

\vspace*{0.1cm}
{\Large in the compact lattice $~U(1)~$ theory}

\vspace*{0.8cm}

I.L. Bogolubsky$^a$, V.K.~Mitrjushkin$^a$,
M. M\"uller--Preussker$^b$ and P. Peter$^b$

\vspace*{0.3cm}

\end{center}

{\sl \noindent
 \hspace*{6mm} $^a$ Joint Institute for Nuclear Research, 141980 Dubna, 
                    Russia   \\
 \hspace*{6mm} $^b$ Humboldt-Universit\"at zu Berlin, Institut f\"ur Physik, 
		    D-10115 Berlin, Germany}

\vspace{0.5cm}
\begin{center}

\renewcommand{\thefootnote}{\fnsymbol{footnote}}
\setcounter{footnote}{0}

\vspace{1cm}
{\bf Abstract}
\end{center}

The Gribov ambiguity problem is studied for compact $U(1)$ lattice theory
within the Lorentz gauge. In the Coulomb phase, it is shown that apart
from double Dirac sheets all gauge (i.e. Gribov) copies originate 
mainly from the zero-momentum modes of the gauge fields. The removal 
of the zero--momentum modes turns out to be necessary for reaching the 
absolute maximum of the gauge functional $~F(\theta)~$. 
A new gauge fixing procedure -- zero-momentum Lorentz gauge -- is
proposed.

\section{Introduction} \setcounter{equation}{0}

To gain a better understanding of the structure of the lattice theory 
and to interprete correctly the numbers obtained in Monte Carlo simulations, 
it is very instructive to compare gauge variant quantities such 
as gauge and fermion field propagators with the corresponding analytical 
perturbative results. In this respect, compact $~U(1)~$ pure gauge theory 
within the Coulomb phase serves as a very useful `test ground', because in 
the weak coupling limit this theory is supposed to describe noninteracting 
photons.  However, previous lattice studies \cite{nak1,bmmp,clas,zmod,bddm} 
have revealed some rather nontrivial effects. It has been shown that the 
standard Lorentz (or Landau) gauge fixing procedure leads to a 
$~\tau$--dependence of the transverse gauge field correlator 
$~\Gamma_T(\tau;\vp)~$ being inconsistent with the expected zero-mass 
behavior \cite{nak1}. Numerical \cite{bmmp,bddm,for} and analytical 
\cite{clas} studies have shown that there is a connection between `bad' 
gauge (or Gribov) copies and the appearance of periodically closed double 
Dirac sheets (DDS).  The removal of DDS restores the correct perturbative 
behavior of the transverse photon correlator $~\Gamma_T(\tau;\vp)~$ with 
momentum $~\vp\ne 0~$. However, it does not resolve the Gribov ambiguity 
problem \cite{grib} completely. 
Other Gribov copies connected with zero--momentum modes 
of the gauge fields still appear, which can `damage' such observables as 
the zero--momentum gauge field correlator $~\Gamma(\tau;0)~$ or the 
fermion propagator $~\Gamma_{\psi}(\tau)~$ \cite{zmod,bddm,zmlf}.

After having understood, why gauge variant lattice correlators behave 
unexpectedly from the perturbation theory point of view, one can search 
for the `true' gauge fixing procedure. This constitutes the main goal of 
this note. We propose a {\it zero--momentum Lorentz gauge} (ZML), which 
permits to get rid of the lattice artifacts and provides correct values for 
various correlation functions.

We are going to compare also the standard Lorentz gauge fixing procedure (LG)
\cite{mo1} with the axial Lorentz gauge (ALG) proposed in \cite{nak2}. We
show that ALG produces just the same problems as LG does and,
therefore, cannot resolve the Gribov ambiguity problem.

\section{Gauge fixing : Lorentz gauge and axial Lorentz gauge}
\setcounter{equation}{0}

The standard Wilson action with $~U(1)~$ gauge group is \cite{wil}

\eq
S(U ) = \beta \sum_x \sum_{\mu > \nu }
         \Bigl( 1 - \cos \theta_{x,\mu\nu} \Bigr) ~,
                      \label{action_u1}
\en

\noi where the link variables are 
$~U_{x\mu} =\exp (i\theta_{x\mu}) \in U(1)~$ and
$~\theta_{x \mu} \in (-\pi, \pi]~$.  The plaquette angles are given by
$~\theta_{x;\mu\nu} = \theta_{x;\mu} + \theta_{x+{\hat \mu};\nu}
- \theta_{x+{\hat \nu};\mu} - \theta_{x\nu}~$. 
This action  makes the part of the full QED action $S_{QED}$, which is
supposed to be compact if we consider QED as arising from a subgroup of
a non--abelian (e.g., grand unified) gauge theory \cite{pol2}.

The plaquette angle $~\theta_P \equiv \theta_{x;\, \mu \nu}~$
can be split up: $~\theta_P = [\theta_P] + 2\pi n_P$, where
$~[\theta_{P}] \in (-\pi ;\pi ]$ and $n_P = 0, \pm 1, \pm 2$.
The plaquettes with $~n_P \neq 0~$ are called Dirac plaquettes.
The dual integer valued plaquettes $~m_{x,\mu \nu} =
\frac {1}{2} \varepsilon_{\mu \nu \rho \sigma} n_{x,\rho \sigma}~$
form Dirac sheets \cite{dgt}.
In our calculations we monitored the total number of Dirac plaquettes
$N_{DP}^{(\mu\nu)}$ for every plane $(\mu;\nu)=(1;1),\ldots ,(3;4)$
and $~ N_{DP} \equiv \max_{(\mu\nu)} N_{DP}^{(\mu\nu)}$.
The appearance of periodically closed double Dirac sheet means
that at least for one out of six planes $(\mu\nu)$ the number of Dirac
plaquettes $N_{DP}^{(\mu\nu)}$ should be

\eq
N_{DP}^{(\mu\nu)} \ge 2\frac{V_4}{N_{\mu}N_{\nu}}~.
\en

\noi For example, on the lattice $~12\times 6^3~$ the appearance of DDS
means $~N_{DP} \ge 72$.

In lattice calculations the usual choice of the Lorentz (or Landau)
gauge is

\eq
\sum_{\mu=1}^4 \bpartial_{\mu} \sin \theta_{x\mu} = 0,
                                                  \label{gf1}
\en

\noi which is equivalent to finding an extremum of the 
functional $F(\theta)$

\eq
F(\theta) = \frac{1}{V_4} \sum_{x} F_{x}(\theta )~;
\qquad F_{x}(\theta) = \frac{1}{8}\sum_{\mu=1}^4
\Bigl[ \cos \theta_{x\mu}+ \cos \theta_{x-{\hat \mu};\mu}\Bigr]
                                                 \label{gf2}
\en

\noi with respect to the (local) gauge transformations

\eq
U_{x\mu} \lra \Lambda_{x} U_{x \mu} \Lambda_{x+{\hat \mu}}^{\ast}~;
~~\Lambda_x=\exp\{i\Omega_x\} \in U(1)~. 
                                 \label{gauge_tr}
\en

\begin{figure}[pt]
\vspace{14.0cm}
\includegraphics{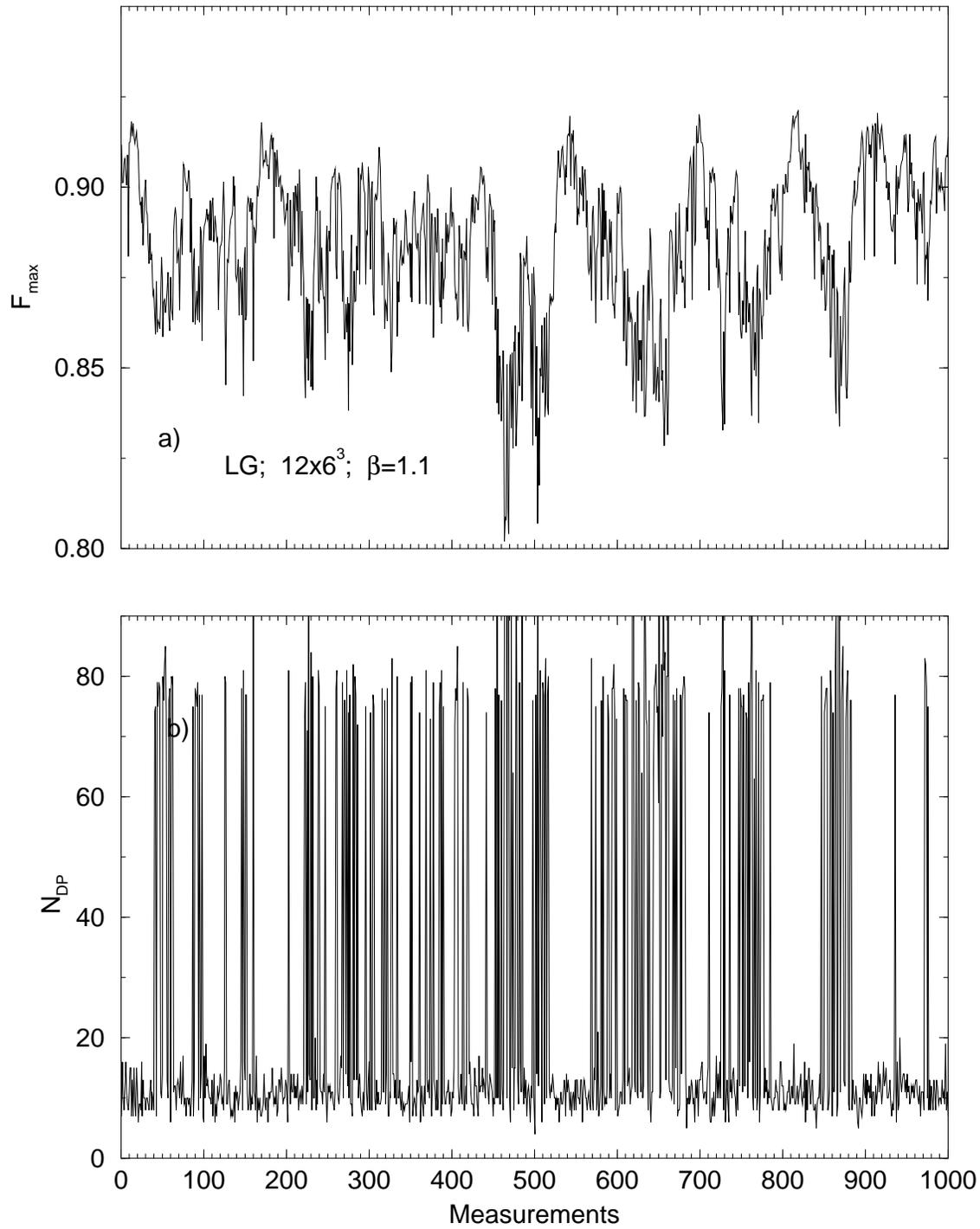}
\vspace{1.0cm}
\caption{Time history of $F_{max}$ ({\bf a}) and $~N_{DP}$
({\bf b}) at $\beta=1.1$ on the $12\cdot 6^3$ lattice in
the standard Lorentz gauge. 
}
\label{fig:fmax_12x06_b01p10_LG}
\end{figure}

\begin{figure}[pt]
\vspace{14.0cm}
\includegraphics{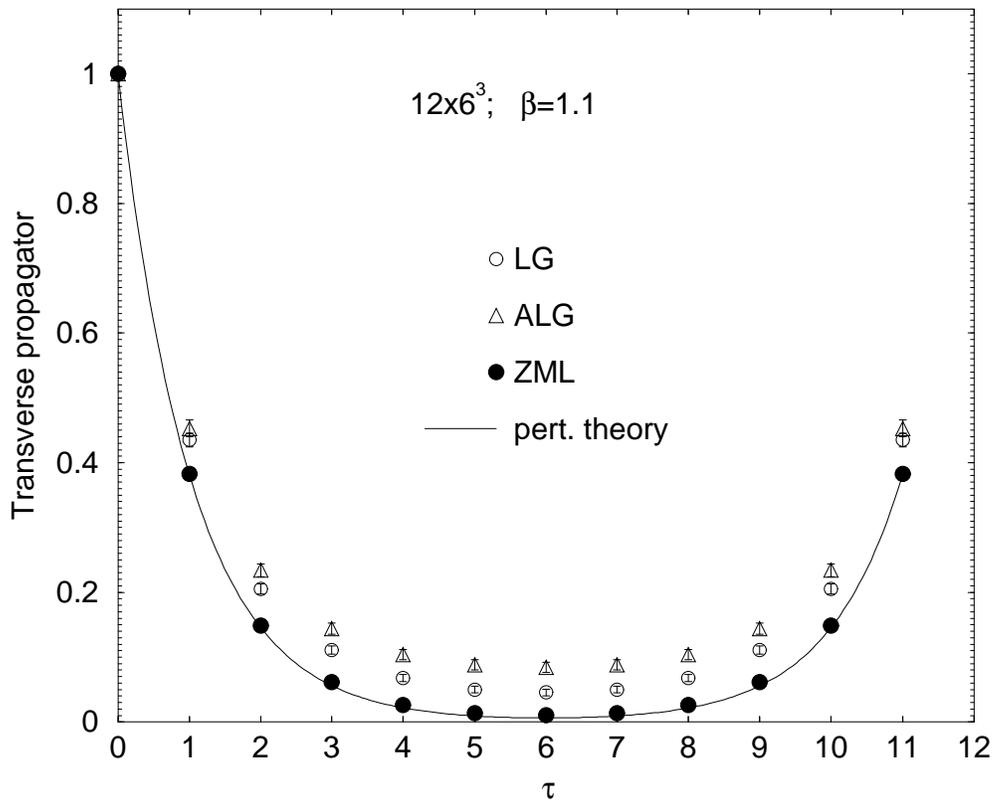}
\vspace{1.0cm}
\caption{Transverse propagator at $\beta=1.1$ on the $12\cdot 6^3$ lattice 
in three different gauges.
}
\label{fig:pcr_12x06_b01p10_3gauge}
\end{figure}

The standard gauge fixing procedure (referred in what follows as LG)
consists of the maximization of the value $~F_x(\theta)~$ at some site
$~x~$ under the local gauge transformations $~\Lambda_x~$, then at
another site and so on. After a certain number of gauge fixing sweeps a
local maximum $~F_{max}~$ of the functional $~F(\theta^{\Lambda})~$ 
is reached. 
In order to improve the convergence of the iterative gauge fixing procedure
one can use the optimized overrelaxation procedure \cite{mo3} 
with some parameter $\alpha$ 
which depends on the volume and $~\beta~$.
For convergence criteria we use 
$$ \max \left |\sum_{\mu} \bpartial_\mu\sin \theta_{x\mu} \right |< 10^{-5} 
\qquad \mbox{and} \qquad
\frac{1}{V} \sum_x \left|\sum_\mu \bpartial_\mu\sin \theta_{x\mu}
\right |< 10^{-6}~.$$
\noi In Figure \ref{fig:fmax_12x06_b01p10_LG}a we show
a typical Monte Carlo time history of $~F_{max}~$ 
obtained using the standard Lorentz gauge fixing procedure.
Partially, the comparatively big dispersion of $~F_{max}~$
is due to periodically closed double Dirac sheets \cite{bmmp,clas}.
In Figure \ref{fig:fmax_12x06_b01p10_LG}b we show the corresponding
time history of $~ N_{DP} = \max_{(\mu\nu)} N_{DP}^{(\mu\nu)}~$. 
For the given lattice size and $\beta$--value $~\sim 20\%~$ 
of configurations turn out to possess DDS. Exactly because of the 
configurations containing DDS the transverse photon correlator 

\eq
 \Gamma_T(\tau;\vp)~=~\langle \Phi(\tau;\vp) ~\Phi^{\ast}(0;\vp) \rangle ~, 
\qquad \Phi(\tau;\vp)~=~
 \sum_{\vx} \exp(i \vp \vx +\frac{i}{2}p_{\mu}) ~\sin \theta_{\tau \vx, \mu}
\en
\noi $(~\mu=1,3, ~~\vp=(0,p,0)~) $ exhibits an unphysical `tachion--like'
behavior \cite{bmmp,clas,bddm}. In Figure
\ref{fig:pcr_12x06_b01p10_3gauge} we show the 
normalized photon correlator $~\Gamma_T(\tau;\vp) / \Gamma_T(0;\vp)~$
for lowest non-vanishing momentum and LG together with results obtained with
other Lorentz gauge fixing procedures to be discussed lateron.  We
clearly see for LG the deviation from the expected zero-mass behavior.
All the observations described above have not been seen changing, when
$~\beta~$ and/or the lattice size were increased considerably
\cite{bddm}.

It is a long--standing believe \cite{zwan} that the
`true' gauge copy corresponds to the {\it absolute} maximum of $~F(\theta)$.
Thus it would be highly desirable to find a Lorentz gauge fixing
procedure which produces a (unique) gauge copy of the given configuration 
with the absolute maximum of $~F(\theta)~$.
However, if one repeatedly subjects a configuration
$~\bigl\{\theta_{x\mu}\bigr\}~$ to a random gauge transformation as in
Eq.(\ref{gauge_tr}) and then subsequently applies to it the LG
procedure, one usually obtains gauge (Gribov) copies with  different
values of $~F_{max}$.

An attempt to resolve this problem has been made in \cite{nak2} where
a modified Lorentz gauge fixing procedure has been proposed. This 
procedure (which we refer to as the axial Lorentz gauge fixing or ALG)
consists of the two steps:

\begin{itemize}
\item[i)] first transform every configuration to satisfy a maximal tree
temporal gauge condition (`axial' gauge) \cite{creu}; 
\item[ii)] then apply the Lorentz gauge
fixing procedure.\footnote{In \cite{nak2} an analytically known 
local solution of the Lorentz gauge condition has been applied, instead
of an iteratively obtained one. But, as we have convinced ourselves, this
solution does not resolve the problems we are discussing here.}    

\end{itemize}

\noi 
An axial gauge with a chosen maximal tree is unique by definition.  
In practice, this is easily checked by random gauge
transformations applied first. Consecutive gauge fixing steps 
-- e.g. the Lorentz gauge iterations --
will lead always to the same result as long as we do not change the
detailed prescription for these steps. The question is, whether this
'unique' Lorentz gauge obtainable for each gauge field configuration 
resolves the problems mentioned above. 
The answer is 'no'. We do not find the absolute maximum of the gauge 
functional in the majority of the cases. There is a quite high percentage 
of Gribov copies left containing DDS (around $~10 \%~$ for $~\beta = 1.1~$ 
and a $~12\times 6^3~$ lattice). As a consequence the transverse non-zero 
momentum photon propagator does not come out correctly again. The 
corresponding data are shown in Figure \ref{fig:pcr_12x06_b01p10_3gauge}, too. 
Doubling of the linear lattice size and enlarging $~\beta~$ (we checked 
$~\beta=2.0~$) do not improve the behavior.

In paper \cite{bmmp} a Lorentz gauge fixing prescription 
with a preconditioning step based on a non-periodic gauge transformation
was proposed. The latter has been chosen in such a way that 
the spatial Polyakov loop averages were transformed into real numbers
as a first step. We convinced ourselves that the Lorentz gauge fixed
configurations with very high probability did not contain DDS. 
As a consequence, the photon correlator becomes correct.
However, this gauge fixing procedure does not provide the absolute 
maximum of the Lorentz gauge functional, too.
Thus, the lattice Gribov problem for QED in the Coulomb phase has not 
been solved.

\section{Zero--momentum Lorentz gauge (ZML)}
\setcounter{equation}{0}

As is well known, apart from the local symmetry Wilson action $~S~$ has an 
additional (global) symmetry with respect to non--periodic transformations

\eq
U_{x\mu} \lra {\bar \Lambda}_x U_{x\mu} {\bar\Lambda}^{\ast}_{x+{\hat \mu}}
= U_{x\mu} \cdot e^{-ic_{\mu}}~;
\qquad
{\bar\Lambda}_x = e^{i\sum_{\mu}c_{\mu}x_{\mu}}~,
                   \label{global1}
\en
\noi or equivalently

\eq
\theta_{x\mu} \lra \theta_{x\mu} -c_{\mu}~.
               \label{global2}
\en

\noi  Note that these transformations do not spoil the periodicity of the 
gauge fields $~U_{x\mu}~$ and $~\theta_{x\mu}~$,
respectively. The transformation in 
Eq.(\ref{global2}) changes the zero--momentum mode $~\phi_{\mu}~$ of the 
link angle $~\theta_{x\mu} \equiv \phi_{\mu} + \delta\theta_{x\mu}~$, where 
$~\sum_x \delta\theta_{x\mu}=0~$, and a proper choice of the parameters 
$~c_{\mu}~$ can make $~\phi_{\mu}~$ equal to zero.

It is rather evident that for the infinitesimal fluctuations 
$~\delta\theta_{x\mu}~$ about the zero--momentum mode $~\phi_{\mu}~$
the absolute maximum of the functional $~F(\theta)~$ corresponds to
the case $~\phi_{\mu}=0~$.

Our main statement (to be proved in this section) is that in most of
the cases ($\ageq 99.99\%$) gauge copies of the given configuration
$~\theta_{x\mu}~$ are due to $~a)$ double Dirac sheets; $~b)$ 
the zero--momentum modes of this field.
It is the exclusion of DDS and of the zero--momentum modes that 
permits to obtain a gauge copy of the given configuration with the 
{\it absolute} maximum of the functional $~F(\theta)~$.

We define our gauge fixing prescription (which we refer to as the 
zero--momentum Lorentz gauge or ZML gauge) as follows.  Every iteration 
consists of one sweep with (global) transformations as in 
Eq.(\ref{global2}) and one sweep with local 
gauge transformations as in the standard Lorentz gauge fixing procedure. 

\begin{figure}[pt]
\vspace{14.0cm}
\includegraphics{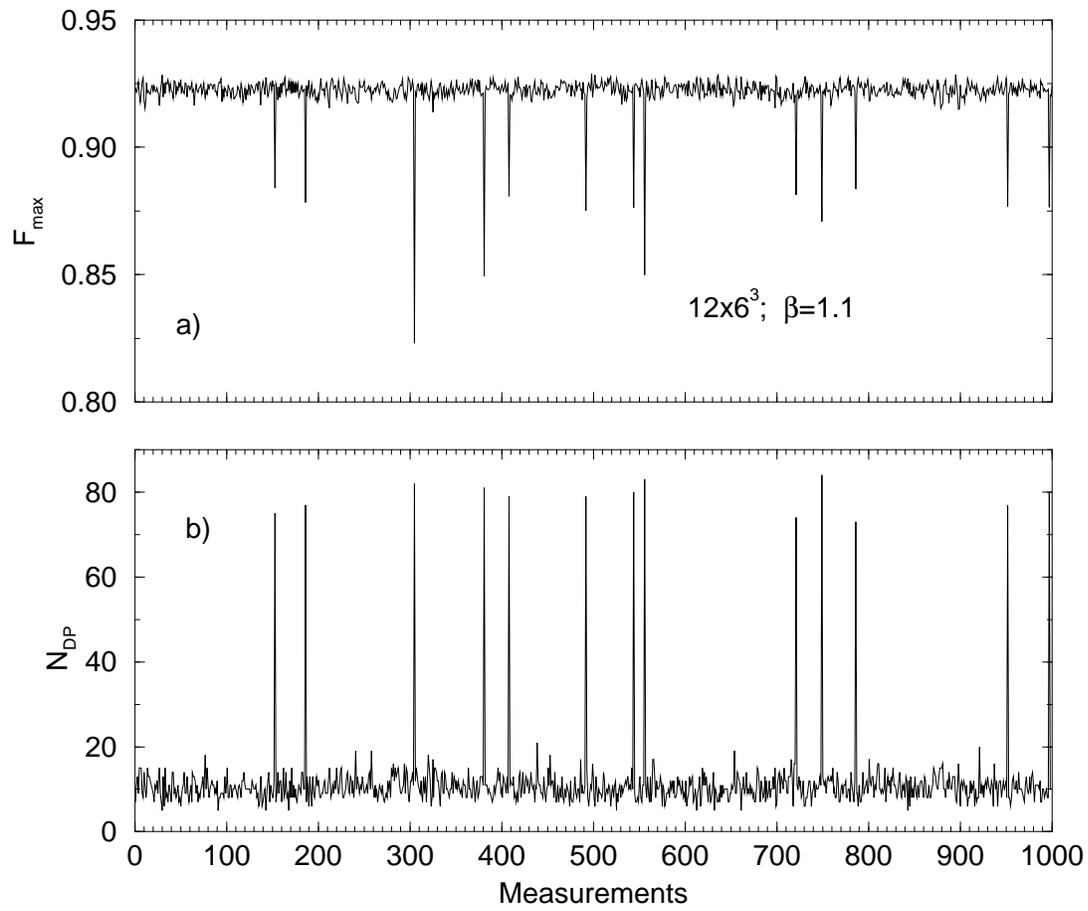}
\vspace{1.0cm}
\caption{Time history of $F_{max}$ ({\bf a}) and $~N_{DP}$
({\bf b}) at $\beta=1.1$ on the $12\cdot 6^3$ lattice in
ZML gauge. 
}
\label{fig:fmax_12x06_b01p10_ZML}
\end{figure}

In Figures \ref{fig:fmax_12x06_b01p10_ZML}a,b we show time histories of 
$~F_{max}~$ and of $~N_{DP}~$ for the gauge fixing
procedure with the minimization of the zero--momentum modes of the gauge 
field. Both the time histories should be compared with the corresponding 
ones for the standard Lorentz gauge shown in Figures
\ref{fig:fmax_12x06_b01p10_LG}a,b. One can see that after  
suppression of the zero--momentum modes the average value of 
$~F_{max}~$ became essentially larger as compared 
with the LG case. At the same time the number of DDS (to be precise, 
the number of configurations with $~N_{DP} \ge 72$) has drastically decreased
($\sim 1\%$). Typically, those configurations with DDS have smaller values
of $~F_{max}~$ than configurations without DDS have.

It is a very easy task to remove the remaining DDS. If a DDS in a ZML gauge
has really appeared, then perform a random gauge transformation to the same
field configuration and repeat the ZML procedure again. As a result
gauge field configurations do not contain DDS.

To convince ourselves that the ZML gauge fixing procedure provides an
absolute maximum of the functional $~F(\theta)~$ we generated 
many random gauge copies for every thermalized configuration
$~\{\theta_{x\mu}\}~$. The number of these random gauge copies
$~N_{RC}~$ varied between $~10~$ and $~1000~$ for different
$~\beta$'s and lattices.
Let $~\Bigl\{\theta^{(j)}_{x\mu}\Bigr\}~$ be the $~j^{th}~$ gauge copy
of the configuration $~\Bigl\{ \theta_{x\mu} \Bigr\}~$ obtained with
the random gauge transformation as in Eq.(\ref{gauge_tr}), $~j=1,\ldots ,
N_{RC}~$. 
For any configuration $~\Bigl\{ \theta_{x\mu} \Bigr\}~$ we define a 
`variance' $~\delta F_{max}(\theta)~$ of $~F_{max}(\theta)~$

\eq
\delta F_{max}(\theta) = 
\max_{ij}\Bigl( F_{max}(\theta^{(i)}) - F_{max}(\theta^{(j)}) \Bigr)~;
~~ i,j =1,\ldots , N_{RC}~.
\en

\begin{figure}[pt]
\vspace{14.0cm}
\includegraphics{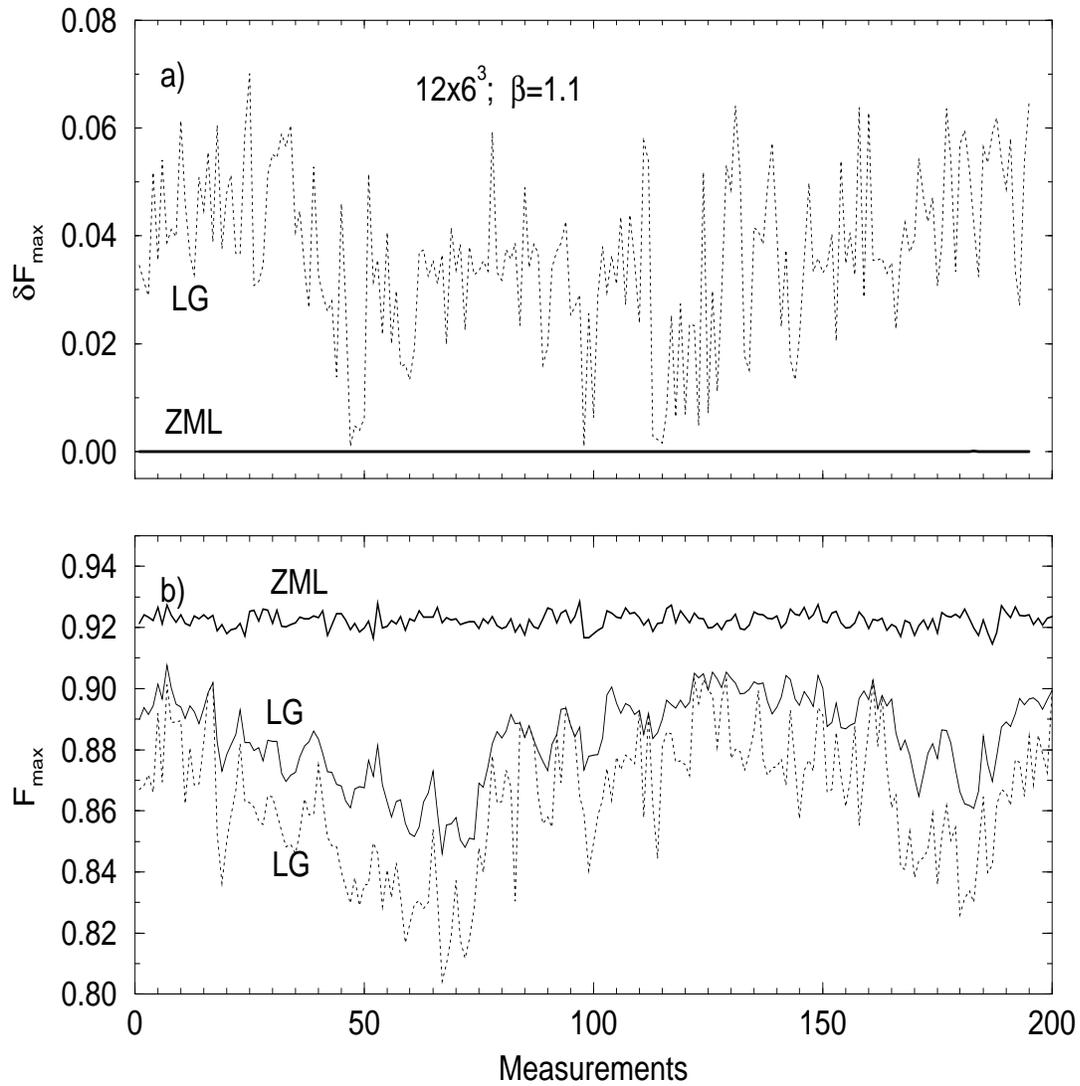}
\vspace{1.0cm}
\caption{Time history of the `variance' $\delta F_{max}$ ({\bf a}) and 
$F_{max}$ ({\bf b}) at $\beta=1.1$ on the 
$12\cdot 6^3$ lattice. LG solid line corresponds to $~N_{RC}=10$. 
Gauge copies with DDS are excluded.
}
\label{fig:var_12x06_b01p10}
\end{figure}

\noi Different gauge copies can, in principle, have different values
of $~F_{max}(\theta)~$. Therefore, its `variance'
  $~\delta F_{max}(\theta)~$ can be non--zero.
For example, the standard gauge fixing procedure (LG) described above 
gives typical values $~\delta F_{max}(\theta)\sim 0.06\div 0.07~$
for  $~N_{RC}~=~10~$.
As an example we show in Figure \ref{fig:var_12x06_b01p10}a a time
history of the `variance' $~\delta F_{max}(\theta)~$ for the standard 
LG and for ZML. In the case of ZML this `variance' is {\it zero}
\footnote{In rather rare cases ($\aleq 0.01\%$) in ZML gauge $~\delta 
F_{max}(\theta)\ne 0~$. So far we have no explanation for it.}.

In Figure \ref{fig:var_12x06_b01p10}b we compare values of $~F_{max}~$
for these two gauges. The lower broken line for the LG corresponds to
the first gauge copy which is just a thermalized configuration produced
by our updating subroutine. Applying random gauge transformations we
produced different random gauge copies ($N_{RC}=10$ in this case)
and then chose the gauge copy with the maximal value of $~F_{max}~$. The
corresponding values are represented by the solid LG line in Figure
\ref{fig:var_12x06_b01p10}b. One can see that even this is far below the
solid line which corresponds to the ZML gauge. Increasing the number
$N_{RC}$ up to $1000$ does not change this conclusion.

One reason for the non--zero value of the `variance' is the appearance of
gauge copies containing double Dirac sheets  \cite{bmmp,clas,bddm}.
Gauge copies with double Dirac sheets have typically much lower values
of $~F_{max}(\theta)~$ as compared to $~F_{max}~$ for gauge copies
without DDS.  Moreover, a contribution of configurations with DDS
`spoils' the photon correlator $~\Gamma_T(\tau;{\vec p})~$ with $~{\vec
p}\ne 0~$ which leads to a wrong dispersion relation inconsistent with
the dispersion relation for the massless photon.  Double Dirac sheets
represent a spectacular example of lattice artifacts which can lead to
a misleading interpretation of the results of numerical calculations
(see also \cite{neu}).

But, as we have seen, the mere exclusion of gauge copies with DDS in the 
standard Lorentz gauge fixing procedure does {\it not} yet provide a 
zero value of the `variance' $~\delta F_{max}(\theta)~$. 
Different gauge copies (without DDS) still have different values
of $~F_{max}(\theta)~$. The exclusion of the zero-momentum modes turns
out to be sufficient to remove the ambiguity.

\section{Conclusions} \setcounter{equation}{0}

Now let us summarize our findings.

\vspace{0.5cm}
Our main result is that $~\sim 99.99\%~$ (if not all) gauge copies for the
given gauge field configuration are due to two reasons :

\begin{itemize}

\item[a)] periodically closed double Dirac sheets;

\item[2)] 
the zero--momentum mode of the gauge field
$~\theta_{x\mu}~$.

\end{itemize}

\noi We didn't find any other reason for the appearance of the gauge
copies. 

\noi The minimization of the zero--momentum mode can be performed 
sweep by sweep using a global transformation as in Eq.(\ref{global2}). 
We proposed a modified gauge fixing procedure consisting of local
gauge transformations in Eq.(\ref{gauge_tr}) and global transformations
in Eq.(\ref{global2}) (ZML gauge).
The exclusion of the DDS (which appear rather rarely in the ZML gauge) can 
be easily performed on the algorithmic level as described in the text.

The application of the ZML gauge fixing procedure provides us with 
{\it absolute} maximum of the functional $~F(\theta)~$.

We have shown that the gauge fixing procedure with axial gauge
preconditioning (ALG) cannot solve the problem of the Gribov ambiguity
in this theory.  The axial gauge preconditioning cannot exclude the
appearance of DDS as well as of nonzero values of the zero--momentum
mode.

In this paper we present our results only for the bosonic sector
of the theory. However, we believe that this study solves ultimatively
the Gribov ambiguity puzzle in the case of quenched compact QED within
the gauge as well as the fermionic sector.
Work on the fermion case is in progress \cite{zmlf}.

The inclusion of dynamical fermions changes somewhat the symmetry group
of the full action, i.e. the parameters $~c_{\mu}~$ in Eq.'s 
~(\ref{global1}),~(\ref{global2}) can have only discrete values.
This case needs some additional study.

\section*{Acknowledgements} \setcounter{equation}{0}

Financial support from grant INTAS-96-370, RFBR grant 99-01-01230  
and the JINR Heisenberg-Landau program is kindly acknowledged.

\vspace{0.5cm}
\section*{Figure captions}

\noi Figure 1. ~~
Time history of $F_{max}$ ({\bf a}) and $~N_{DP}$
({\bf b}) at $\beta=1.1$ on the $12\cdot 6^3$ lattice in
the standard Lorentz gauge. 

\vspace{0.25cm}
\noi Figure 2. ~~
Transverse propagator at $\beta=1.1$ on the $12\cdot 6^3$ lattice 
in three different gauges.

\vspace{0.25cm}
\noi Figure 3. ~~
Time history of $F_{max}$ ({\bf a}) and $~N_{DP}$
({\bf b}) at $\beta=1.1$ on the $12\cdot 6^3$ lattice in
ZML gauge.

\vspace{0.25cm}
\noi Figure 4. ~~
Time history of the `variance' $\delta F_{max}$ ({\bf a}) and 
$F_{max}$ ({\bf b}) at $\beta=1.1$ on the 
$12\cdot 6^3$ lattice.
LG solid line corresponds to $~N_{RC}=10$. 
Gauge copies with DDS are excluded.

\newpage

\end{document}